# Impact of COVID-19 on Exchange rate volatility of Bangladesh: Evidence through GARCH model

**Rizwanul Karim[1]**

**ABSTRACT**

This study uses the GARCH (1,1) model to examine the impact of COVID-19 cases (log value) on the volatility of the Exchange rate return of Bangladeshi taka (BDT) over the US dollar (USD), Japanese Yen (JPY), and Swedish Krona (SEK). The result shows that an increase in the number of COVID-19-affected cases in Bangladesh has a significant and positive impact on the volatility of exchange rates BDT/USD, BDT/JPY, and BDT/SEK.

**Keywords:** "COVID-19", "Exchange rate", "GARCH", "Volatility"

[1] Joint Director, Banking Regulations and Policy Department, Bangladesh Bank

# Chapter One: Introduction

## 1.1 Significance

### 1.1.1 Background

The coronavirus (COVID-19) pandemic started at the end of December 2019. Since outbreak, COVID-19 has prompted a great deal of worldwide economic and financial uncertainty (Zeren and Hizarci, 2020; Khan *et al.,* 2020). The spread of this virus and globally enforced lockdowns have negatively impacted the overall demand and created significant short-term food price volatility (Albulescu, 2020). The fear stimulated by the virus has put substantial stress on financial markets, where price volatility has been increasing continuously. The lockdown restrictions widely implemented across the globe to curb the spread of the virus including travel prohibitions and border closures, stay-at-home and work-from-home orders, and extensive business closures, all resulted immense fallout for the global economy.

In this scenario, the negative economic impact of this disease on economy is imminent. In case of Bangladesh as the export hampered, the Bangladeshi taka (BDT) exchange rate is expected to be disturbed too. Due to countrywide lockdown the movement of people and goods to and from this place was completely stopped. Due to the novel nature of COVID-19, it became difficult to ascertain if the virus could be spread through goods transport or not. Production had already suffered due to complete lockdown in the country and then reduced demand overseas added further to the declining exports. All these factors can affect the value of BDT. In such a scenario, it is interesting to know how the BDT exchange rate moved with the emerging situation of the COVID-19 outbreak, explicitly speaking the number of daily new confirmed cases in Bangladesh during this period.

Motivated with this background, we draw the attention to the COVID-19 effects in the context of exchange rate of BDT. This study is the first attempt to explore if the covid-19 drives exchange rate volatility of BDT. This paper estimates the impact of daily reported Covid-19 cases on both the conditional mean and the conditional volatility of the returns of BDT exchange rate. The results reported are based on data for March 09, 2020 to December 30, 2020 and consider the impact of reported cases on BDT/USD, BDT/JPY and BDT/SEK volatility.

## 1.2 Literature review

Because of COVID-19 whole world is suffering from massive economic damages. The crisis spilled over and became the catalyst for a global financial crisis. Even before COVID-19, financial institutions experienced extended competition due to worldwide financial collapse as well as new regulations (Anjom & Faruq, 2023). In this context, a number of researches are ongoing to measure the economic and social impact of the crisis caused by the pandemic. Goodell (2020) presents Agendas for future research about the impact of Covid-19 on financial markets and institutions. Albulescu (2020) explores the impact of Covid-19 numbers on crude oil prices, while controlling for the impact of financial volatility and the United States (US) economic policy uncertainty. He shows a marginal negative impact of Covid-19 on the crude oil prices in the long run. Zhang et al. (2020) examine the impacts of coronavirus on global financial markets and show substantial increases of volatility in global markets due to the outbreak. They find that global stock markets linkages display clear different patterns before and after the pandemic announcement and policy responses introduce further uncertainties in the global financial markets. Moreover, Al-Awadhi et al. (2020) investigate whether contagious infectious diseases affect Chinese stock market outcomes and find that Covid-19 has significant negative effects on stock returns. Lately, Onali (2020) finds that an increase in the reported deaths in the US has a positive impact on the conditional heteroscedasticity of the Dow Jones Index.

Since Engle introduced autoregressive conditional heteroscedasticity (ARCH) models in 1982, this model, along with the Generalized autoregressive conditional heteroscedasticity (GARCH) model (Bollerslev,1986), become standard tools for assessing the volatility of financial variables. According to Sadorsky (2006), GARCH is extremely useful for observing heteroskedastic behavior or volatility clustering in financial markets without requiring higher-order models. This model works by predicting variance in the current period by forming a weighted average of a) a long-term average, b) the variance forecasted in the previous period, and c) information about volatility observed in the previous period. The model also appropriately represents the volatility clustering often witnessed in financial returns data, where significant changes in returns tend to be followed by even more substantial changes (Trück, 2020; Yousef, 2020).

Since the model is very much suitable to examine volatility in financial market, several studies have been done to determine exchange rate volatility using GARCH model. Bala and Asemota (2013) examined exchange rate volatility using GARCH models. They used monthly exchange rate return series for the naira (Nigerian currency) against the US dollar ($), British pound, and euro. Clement and Samuel (2011) also aimed to model Nigerian exchange rate volatility. They used the monthly exchange rate of the naira against the US dollar and British pound for the period from 2007 to 2010. Rofael and Hosni (2015) aimed to forecast and estimate exchange rate volatility in Egypt using ARCH and state space (SS) models. Using daily exchange rate data covering about 10 years, they found volatility clustering in Egypt's exchange rate returns, as well as a risk of mismatch between exchange rates and the stock market. Similar results were obtained by Choo, Loo, and Ahmad (2002), who used GARCH model variants to capture the exchange rate volatility dynamics of the Malaysian ringgit (RM) against the pound sterling. They used daily data for the period from 1990 to 1997 and concluded that the volatility of the RM–sterling exchange rate was persistent.

In case of BDT, Abdullah, Siddiqua, Siddiquee and Hossain (2017) examined the volatility of Exchange rate of BDT/USD through GARCH. They used daily exchange rate return for BDT/USD. Still there is no published study which examines the volatility of exchange rate as an impact of COVID-19 outbreak. This study is the first attempt to examine the impact of COVID-19 outbreak on volatility of exchange rate return of BDT.

# Chapter Two: COVID-19 outbreak in Bangladesh

## 2.1 COVID-19 scenario in Bangladesh

The COVID-19 pandemic in Bangladesh is part of the worldwide pandemic of coronavirus disease 2019. The virus stared to spread in Bangladesh from March 2020. This pandemic significantly changed the worldwide scenario in many aspects (Raihan et al, 2023). The first three known cases were reported on March, 8 2020 by the country's epidemiology institute, IEDCR. Since then, the pandemic has spread day by day over the whole nation and the number of affected people has been growing.

In order to control the spread of COVID-19 the government declared "lockdown" throughout the nation from 23 March to 30 May. Infections remained low until the end of March but saw a steep rise in April. In the week ending on 11 April, new cases in Bangladesh increased by 1,155 percent, the highest in Asia. Bangladesh reached affected case of 160,000 and death cases of 2,000 on July 5, 2020.

Figure:1 depicts the number of COVID-19 affected, death and recovery rate in Bangladesh.

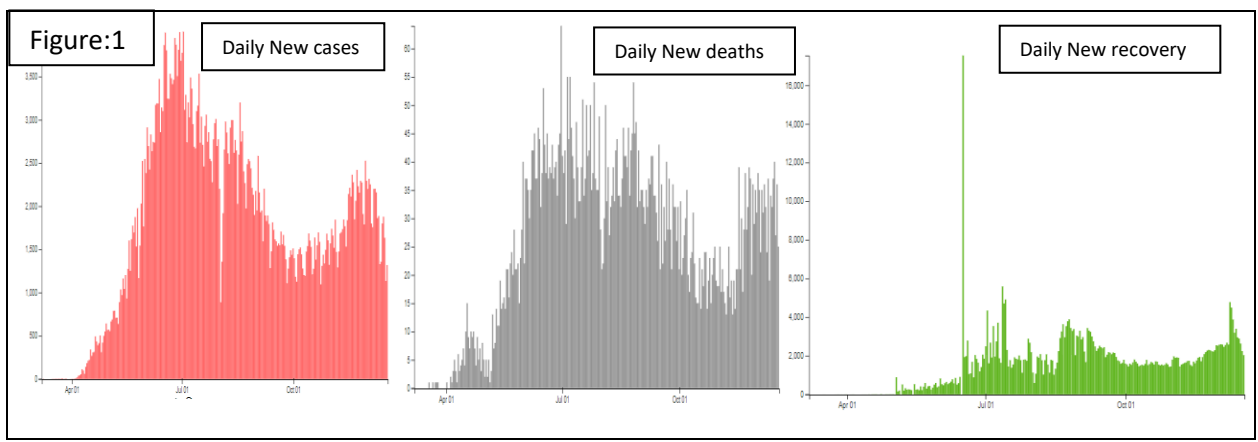

Source: https://corona.gov.bd/

# Chapter Three: Impact of COVID-19 on volatility of exchange rate return; Evidence through GARCH model

## 3.1 Data and Methodology

### 3.1.1 Data

The study involves calculating the impact of daily COVID-19 affected rate in Bangladesh on volatility of exchange rate return of BDT. Daily exchange rate data is the daily observations of nominal exchange rate of Bangladeshi taka (BDT) supplied by Bangladesh Bank the central bank of Bangladesh. Bangladesh Bank publishes daily buying and selling rate of BDT with other top currencies. For this study average of daily buying and selling rate is taken as the nominal exchange rate. Three currencies US Dollar (USD), Japanese Yean (JPY) and Swedish Krona (SEK) have been considered for the study. Because the nominal exchange rate series is usually nonstationary, it is not appropriate for analysis. Hence the nominal exchange rate is converted into exchange rate return as following logarithmic transformation. In particular, the following formula is used to calculate the rate of return on exchange rate:

$r_t = \ln(fx_t) - \ln(fx_{t-1})$

or, $r_t = \ln(\frac{fxt}{fxt-1})$

Here, $r_t$ stands for exchange rate return at period t. $fx_t$ and $fx_{t-1}$ denote the nominal exchange rate of the BDT at period t and (t–1).

The daily data on the number of Covid-19 reported cases is obtained from the COVID-19 tracker website of Bangladesh. In Bangladesh, the first case of COVID-19 was officially reported on March 08. This study covers the data from March 09, 2020 to December 30, 2020, which gives total 212 observations. As the number of daily COVID-19 affected case is nonstationary, $1^{st}$ difference of the daily affected case is calculated. The statistical software EViews 11 was used to perform the quantitative exercise.

The descriptive statistics of daily volatility of the BDT/USD, BDT/JPY and BDT/SEK. The measures of Covid-19 are given in Table 1.

Table: 1 Summery Statistics

|  | ERR(BDT/USD) | ERR (BDT/JPY) | ERR (BDT/SEK) | COVID Cases (1st difference) |
|---|---|---|---|---|
| Mean | -8.34e-06 | 7.05e-05 | 0.000626 | 5.8254 |
| Median | 0.00 | 0.00 | 0.000195 | 3.00 |
| Maximum | 0.00088 | 0.02853 | 0.2126 | 763.00 |
| Minimum | -0.001242 | -0.0318 | -0.0362 | -1416 |
| Std. Dev. | 0.00018 | 0.005203 | 0.0065 | 259.27 |
| Skewness | -3.127140 | -1.987 | -0.7696 | -0.8593 |
| Kurtosis | 31.13329 | 21.5948 | 8.8454 | 8.100 |
| Jarque-Bera | 7336.95 | 3193.80 | 322.76 | 255.88 |
| Probability | 0.00 | 0.00 | 0.00 | 0.00 |

The return of exchange rate series is showing negative skewness statistics and high kurtosis which depicts the presence of fat tails and a non-symmetric series. These exhibit one of the important characteristics of leptokurtosis. The non-normality condition is supported by a Jarque-Bera test which shows that the null hypothesis of normality is rejected at the five per cent level of significance.

Figure 2 presents the plot of exchange rate return for BDT/USD, BDT/JPY and BDT/SEK. LGRUSDI stands for log return of exchange rate of BDT over USD, LGRJPYI stands for log return of exchange rate of BDT over JPY and LGRSEKI stands for log return of exchange rate of BDT over Swedish Krona. There are clear trends showing upward movements tend to be followed by upward movements and downward movements also followed by other downward movements. So, there exists clustering volatility in the patterns which is a precondition to calculate GARCH model.

Figure 2: Daily return of exchange rate

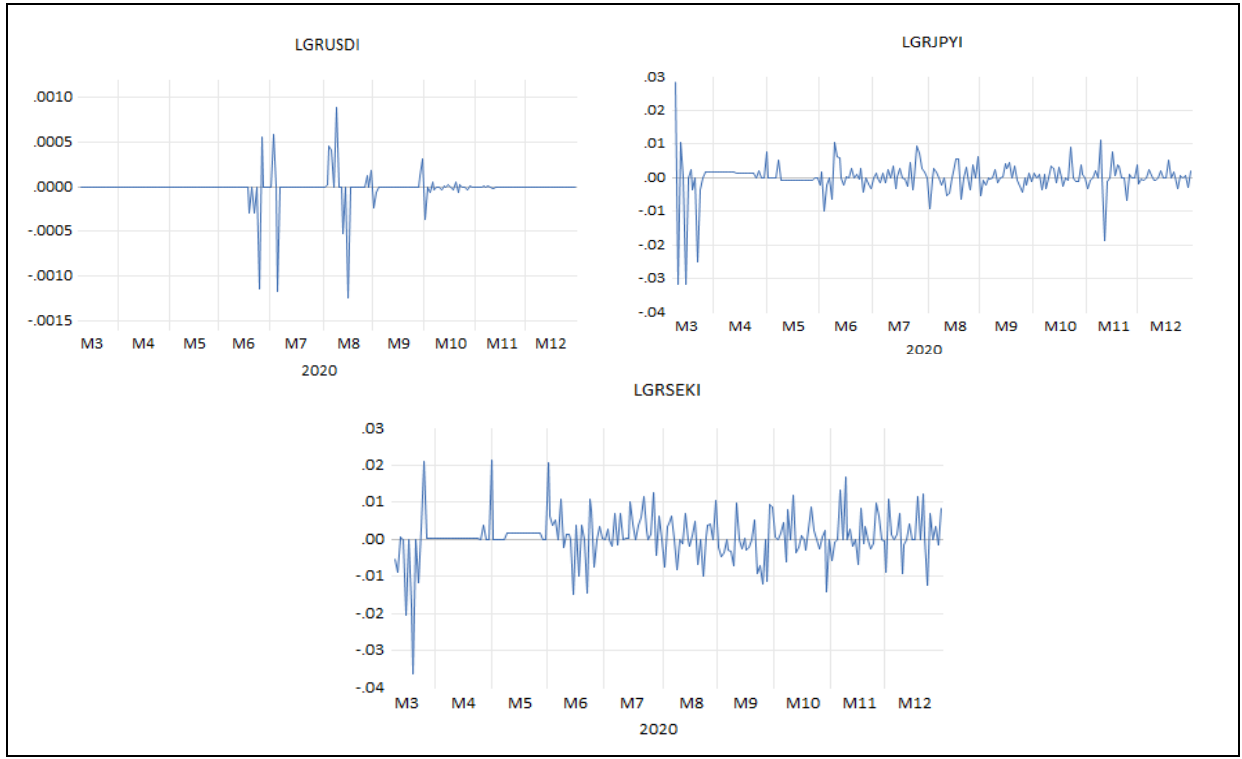

Unit root test: To perform the study is required that the data should be stationary. The ADF and PP test are used to test whether the data is stationary or not. Here the null hypothesis is the data has a unit root. To be nonstationary the null hypothesis has to be rejected. Table 2, shows the result of ADF and PP pest the tests. In all cases the null hypothesis is rejected at 1% level of significance. So, the data is stationary.

Table 2: Unit root test: ADF and PP

| Variables | ADF | PP |
|---|---|---|
| ERR (BDT/USD) | -15.69(0.00*) | -15.65(0.00*) |
| ERR (BDT/JPY) | -18.64(0.00*) | -18.19(0.00*) |
| ERR (BDT/SEK) | -13.88(0.00*) | -14.00(0.00*) |
| COVID Cases (1st deference) | -19.67(0.00*) | -19.48(0.00*) |

*Probability

3.1.2 Methodology

Although there is a variety of different GARCH (Generalized Autoregressive Conditional Heteroskedastic) models, in many empirical applications, the GARCH (1,1) is generally mentioned as the most popular model in the GARCH-type models. Because of its computationally convenience it is widely used by specialists to model volatility of daily returns. It has one ARCH effect and one GARCH effect. This study focuses on the GARCH (1,1) model to analyze the effect of Covid-19 on the volatility of the BDT exchange rate return.

The GARCH (1, 1) model is expressed by the conditional mean equation (eq.1) and the conditional variance equation (eq.2)

Mean equation: $r_t = \alpha_0 + \alpha_1 covid_t + \varepsilon_t$ \hfill (eq.1)

Variance equation: $\sigma_t^2 = h_t = \beta_o + \beta_1 h_{t-1} + \beta_2 u^2_{t-1}$ \hfill (eq.2)

Where: $r_t$ is the return of exchange rate of of BDT/USD, BDT/JPY and BDT/SEK, $covid_t$ is the number of daily Covid affected cases at time t. To make the data of daily affected cases stationary, it is taken as 1st difference form. logarithm function, $\varepsilon_t$ are the innovations. $\sigma_t^2$ is the conditional variance, where $\beta_o$ and $\beta_1$ and $\beta_2$ all are supposed to be >0. Also, if $\beta_1+\beta_2 <1$ The model ensures positive conditional variance stationarity.

## 3.2 Empirical results

The results for the GARCH (1,1) model are reported in Table 4. The results show that both ARCH and GARCH effects are significant. It depicts that number of cases affected by COVID-19 affects the current volatilities of exchange rate return series. Moreover, the results of Table 3 show the mean reverting process because the sum of coefficients is less than one (Coefficient of L-ARCH+ Coefficient of L-GARCH <1), and this is the condition for stationarity or mean reversion (Bollerslev, 1986).

Table 3: Volatility of Exchange rate return, number of affected cases: GARCH (1,1) estimation

| Variables | Vol (BDT/USD) | Vol (BDT/JPY) | Vol (BDT/SEK) |
|---|---|---|---|
| Mean equation: | | | |
| Constant ($\alpha_0$) | 7.43E-07 | 0.000270 | 8.88E-05 |
| Covid$_t$ ($\alpha_1$) | 5.86E-09 (9.43*) | 1.47E-06(2.10*) | 1.14E-06(7.232*) |
| Probability | 0.00 | 0.03 | 0.00 |
| Variance equation: | | | |
| GARCH term ($\beta_1$) | 0.4910(12.5307*) | 0.6726(19.59*) | 0.708(0.652*) |
| ARCH term ($\beta_2$) | 0.2695(0.4620*) | 0.1300(4.46*) | 0.089(0.310*) |
| Constant ($\beta_0$) | 2.76E-11(0.4530*) | 2.09E-06(7.27*) | 3.54E-05(0.258*) |
| Observations | 212 | 212 | 212 |
| Diagnostic test | | | |
| Serial correlation | No | No | No |
| ARCH effect | No | No | No |

* The Z statistics value

The expected result for ($\alpha_1$) in the mean equation is positive. The positive ($\alpha_1$) value indicate that the number of COVID-19 affected cases in Bangladesh impact the volatility of return of exchange rate of BDT. The study also finds the positive vale of ($\alpha_1$) in all three exchange rates BDT/USD, BDT/JPY and BDT/SEK at 1% level of significance. It suggests that there is a positive relationship between the number of COVID-19 affected cases and volatility of exchange rate of return of BDT/USD, BDT/JPY and BDT/SEK and this relationship is statistically significant.

The study also finds that coefficients of the constant variance term GARCH $\beta_1$ and ARCH $\beta_2$ parameters are positive for all the three exchange rates. $\beta_o$, $\beta_1$ and $\beta_2$ all are >1. Besides, $\beta_1+\beta_2$ <1 which fulfills the condition of stationarity or mean reversion.

### 3.3 Robustness check

The robustness of the results is examined to ensure the accuracy of the model. Table 4 shows that there is no serial correlation as the Null hypothesis cannot be rejected at 1% significance level. Beside this, the ARCH effect test suggests that there is no evidence of conditional heteroscedasticity. The residuals are normally distributed, there is no evidence of heteroscedasticity and serial correlation among residuals, indicating a robust model to make unbiased statistical inferences.

### 3.4 Measuring Impact of COVID-19 on ER volatility of BDT by dividing the pandemic period into two parts

To examine the impact of COVID-19 on exchange rate of BDT the COVID-19 period is divided into two parts. One is beginning of the pandemic and the other is during the middle of the pandemic. The beginning of the pandemic time is considered from March 09, 2020 to June 26, 2020. In this part total 80 days daily COVID affected rate and daily nominal exchange rate of BDT is considered. In this study, this period is termed as period 1. On the other hand, the second part is considered from June 29, 2020 to December 30, 2020. In this part total 132 days daily COVID affected data and daily nominal exchange rate data is considered. This period will be termed as period 2. The two periods are divided on the basis of number of affected cases. The early days when the number of affected cases were in rise is termed Period 1, on the other hand after reaching to the peak the number of cases sustained for a prolonged time. This time is treated as period 2.

Then the impact of COVID affected cases on return of Exchange rate of BDT is examined using equation 1 and 2 for both periods separately. Equation 1 is mean model and equation 2 is variance model. The parameters $\alpha_1$, $\beta_1$ and $\beta_2$ is compared for both periods. It is expected that there is change in the parameters value between two periods which indicate the structural break in the GARCH process.

3.4.1 Measuring volatility of exchange rate for Period 1 (Beginning of COVID)

The results for the GARCH (1,1) model of beginning of COVID period are reported in Table 4.

Table 4: Volatility of Exchange rate return, number of affected cases: GARCH (1,1) estimation

| Variables | Vol (BDT/USD) | Vol (BDT/JPY) | Vol (BDT/SEK) |
| --- | --- | --- | --- |
| Mean equation: | | | |
| Constant ($\alpha_0$) | 6.59E-10 (1.70) | 0.000551(1.59) | 0.0013(1.25) |
| Covid$_t$ ($\alpha_1$) | -3.58E-12 (-3.88) | -8.29E-07(-0.47) | -2.20E-06(-0.59) |
| Probability | 0.00 | 0.63 | 0.55 |
| Variance equation: | | | |
| GARCH term ($\beta_1$) | 0.6000(2.03*) | 0.6432(10.81*) | 0.615(5.85*) |
| ARCH term ($\beta_2$) | 0.1500(1.40*) | 0.1723(1.95*) | 0.199(2.88*) |
| Constant ($\beta_0$) | 1.15E-18(1.2669*) | 1.59E-06(6.97*) | 1.01E-05(2.76*) |
| Observations | 80 | 80 | 80 |
| Diagnostic test | | | |
| Serial correlation | No | No | No |
| ARCH effect | No | No | No |

* The Z statistics value

The positive ($α_1$) value indicate that the number of COVID-19 affected cases in Bangladesh impact the volatility of return of exchange rate of BDT. For period 1 (beginning of pandemic) the study finds negative vale of ($α_1$) in all three exchange rates BDT/USD, BDT/JPY and BDT/SEK. The values are not statistically significant also. So, at the beginning at pandemic period, we cannot say that the number of COVID-19 affected cases in Bangladesh impact the volatility of exchange rate of BDT.

The diagnostic test of residual suggests that there is no serial correlation and ARCH effect in the model. This indicates that the model is ok.

3.4.2 Measuring volatility of exchange rate for Period 2 (During COVID)

The results for the GARCH (1,1) model of period 2 are reported in Table 5

Table 5: Volatility of Exchange rate return, number of affected cases: GARCH (1,1) estimation (period 2)

| Variables | Vol (BDT/USD) | Vol (BDT/JPY) | Vol (BDT/SEK) |
|---|---|---|---|
| Mean equation: | | | |
| Constant ($α_0$) | 2.65E-06 (0.57) | 0.000221(0.67) | 0.00016(1.56) |
| Covid$_t$ ($α_1$) | 6.40EE-08 (21.40) | 2.29E-06(2.62) | 1.31E-06(2.23) |
| Probability | 0.00 | 0.008 | 0.02 |
| Variance equation: | | | |
| GARCH term ($β_1$) | 0.40(12.11*) | 0.63(2.75*) | 0.67(1.68*) |
| ARCH term ($β_2$) | 0.109(6.71*) | 0.14(2.255*) | 0.14(1.77*) |
| Constant ($β_0$) | 2.31E-10(6.13*) | 2.66E-06(1.29*) | 1.69E-05(1.02*) |
| Observations | 132 | 132 | 132 |
| Serial correlation | No | No | No |
| ARCH effect | No | No | No |

* The Z statistics value

The expected result for ($\alpha_1$) in the mean equation is positive. The positive ($\alpha_1$) value indicate that the number of COVID-19 affected cases in Bangladesh impact the volatility of return of exchange rate of BDT. For period 2, the study also finds the positive vale of ($\alpha_1$) in all three exchange rates BDT/USD, BDT/JPY and BDT/SEK at 1% level of significance. It suggests that, During Pandemic period, there is a positive relationship between the number of COVID-19 affected cases and volatility of exchange rate of return of BDT/USD, BDT/JPY and BDT/SEK and this relationship is statistically significant.

The study also finds that coefficients of the constant variance term GARCH $\beta_1$ and ARCH $\beta_2$ parameters are positive for all the three exchange rates. $\beta_0$, $\beta_1$ and $\beta_2$ all are >1. Besides, $\beta_1+\beta_2$ <1 which fulfills the condition of stationarity or mean reversion.

The diagnostic test of residual suggests that there is no serial correlation and ARCH effect in the model.

### 3.5 Interpretation of differences of parameters between two periods:

The study finds significant difference in parameters between period 1 and period 2. At the beginning of pandemic all the values of ($\alpha_1$) are found negative. On the contrary, all the values of ($\alpha_1$) are found positive during pandemic period. This result implies that at the beginning of the pandemic the number of COVID-19 affected cases in Bangladesh don't impact the volatility of exchange rate of BDT. On the other hand, during the pandemic period, the number of COVID-19 affected cases impact the volatility of exchange rate of BDT. This result also indicates that there is existence of structural break in the GARCH process. This result supports the fact that during COVID-19 there was stagnant situation in the economic activities. During COVID-19 period, major determinants of exchange rate including export earnings, FDI, current account balance of the country were hampered which facilitate the volatility of exchange rate.

## 3.6 Forecasting future exchange rate volatility

The study uses GARCH model to calculate static and dynamic forecasts of the mean and the conditional variance. To study of the volatility of exchange rate of BDT the study uses 3 forecasting methods as listed below.

1. Root Mean Square Error (RMSE): $\quad RMSE = \sqrt{\frac{1}{n}\sum_{t=1}^{n}(r_t^2 - \sigma_t^2)^2}$ (eq.3)

2. Mean Absolute Error (MAE): $\quad MAE = \frac{1}{n}\sum_{t=1}^{n}|r_t^2 - \sigma_t^2|$ (eq.4)

Here $r_t^2$ is used as a substitute for the realized or actual variance and $\sigma_t^2$ is the forecasted variance.

3. Theil Inequality Coefficient: $\quad U = \dfrac{\sqrt{\frac{1}{n}\sum_{i}(X_i - Y_i)^2}}{\sqrt{\frac{1}{n}\sum_{i}X_i^2} + \sqrt{\frac{1}{n}\sum_{i}Y_i^2}}$ (eq.5)

Thiel's inequality coefficient measures of how well a time series of estimated values compares to a corresponding time series of observed values. It statistically measures the degree at which one time series differs from another one. (vosesoftware.com)

The study examines forecasting future exchange rate volatility of BDT in three parts. The first part forecasts future exchange rate of BDT taking the sample from March 09, 2020 to December 01, 2020 and forecasts volatility of future exchange rate for December 02, 2020 to December 30, 2020. In the second part forecasting of future exchange rate volatility is done for period 1, beginning of the pandemic. In this par the sample size is considered from the period of March 09, 2020 to June 01, 2020, and forecasting is calculated for the period June 02, 2020 to June 26, 2020. Finally, in the last part, forecasting of future exchange rate volatility is done for period 2, during the pandemic time. In this part, the sample size for forecasting is considered from June 29, 2020 to December 01, 2020 and the forecasting is done for the period December 02, 2020 to December 30, 2020. Later, the forecasting result for period 1 and period 2 are compared to find out if there is any improvement in the forecasting of exchange rate volatility.

3.6.1 Forecasting future exchange rate volatility of BDT for whole period:

In this part future exchange rate volatility of BDT is forecasted taking the sample from March 09, 2020 to December 01, 2020 and forecasts volatility of future exchange rate for December 02, 2020 to December 30, 2020. The results are given in table 6.

Table 6: Forecast summery statistics for future exchange rate volatility of BDT (Whole period)

| Volatility of BDT/USD | RMSE | 1.74E-06 |
|---|---|---|
| | MAE | 1.57E-06 |
| | Theil Inequality Coefficient | 1.00 |
| Volatility of BDT/JPY | RMSE | 0.0019 |
| | MAE | 0.0013 |
| | Theil Inequality Coefficient | 0.902 |
| Volatility of BDT/SEK | RMSE | 0.0063 |
| | MAE | 0.0046 |
| | Theil Inequality Coefficient | 0.843 |

The lower the value of RMSE, MAE and Theil Inequality Coefficient the more the model is suitable for forecasting. Results in table 7 show very low value all the parameters MAE, RMSE and Theil Inequality Coefficient. It depicts that the model used in this study (equation 1 and 2) is very much suitable to for forecasting exchange rate of BDT. Figure 3: Future exchange rate volatility forecasting (December 01,2020 to December 30, 2020)

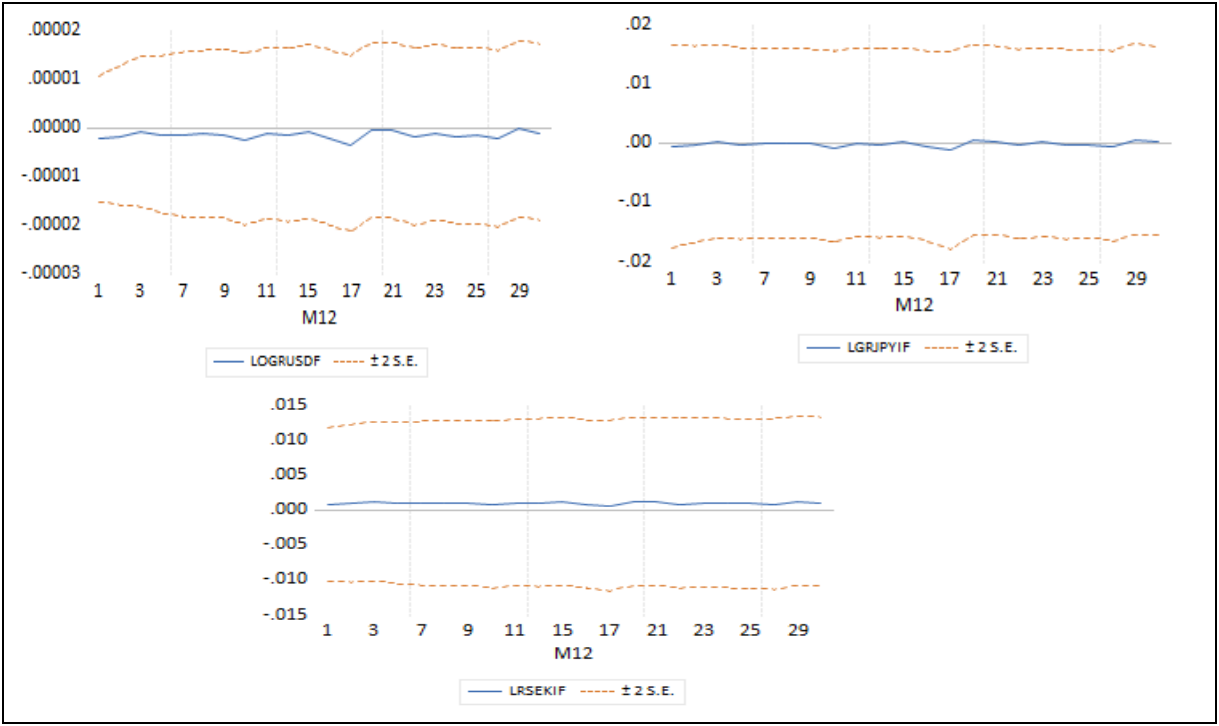

The volatility of exchange rate of BDT for the period from December 01, 2020 to December 30, 2020 is measured through GARCH model. The result is graphically shown in figure 6. LOGRUSDF stands for forecasting of BDT/USD exchange rate, LGJPYIF stands for forecasting of BDT/JPY exchange rate and LGSEKIF stans for forecasting of BDT/SEK exchange rate. Sample of this study covers the period from March 09, 2020 to December 01, 2020 and volatility is measured for the period of one month from December 02, 2020 to December 30, 2020. The result shows that the blue line is very much aligned to gray line and lies between dotted line. It indicates the model forecasted the Exchange rate volatility very well as the original value is close to the forecasted value.

3.6.2 Forecasting future exchange rate volatility for period 1 (beginning of COVID)

In this part future exchange rate volatility of BDT is forecasted taking the sample from March 09, 2020 to June 01, 2020, and forecasting is calculated for the period June 02, 2020 to June 26, 2020. The results are given in table 8

Table 7: Forecast summery statistics for exchange rate volatility of BDT for Phase 1 (beginning of COVID)

| Volatility of BDT/USD | RMSE | 0.000307 |
|---|---|---|
| | MAE | 0.000121 |
| | Theil Inequality Coefficient | 1.00 |
| Volatility of BDT/JPY | RMSE | 0.005 |
| | MAE | 0.003 |
| | Theil Inequality Coefficient | 0.903 |
| Volatility of BDT/SEK | RMSE | 0.007 |
| | MAE | 0.005 |
| | Theil Inequality Coefficient | 0.97 |

Results in table 7 show that almost all the values in the volatility model using total of cases as a proxy of corona pandemic have a low MAE, RMSE and Theil Inequality Coefficient. The lower the value of RMSE, MAE and Theil Inequality Coefficient the more the model is suitable for forecasting. The result of table 7 depicts that the model used in this study (equation 1 and 2) is very much suitable to for forecasting exchange rate of BDT.

Figure 4: Exchange rate volatility forecasting (June 2,2020 to June 26,2020)

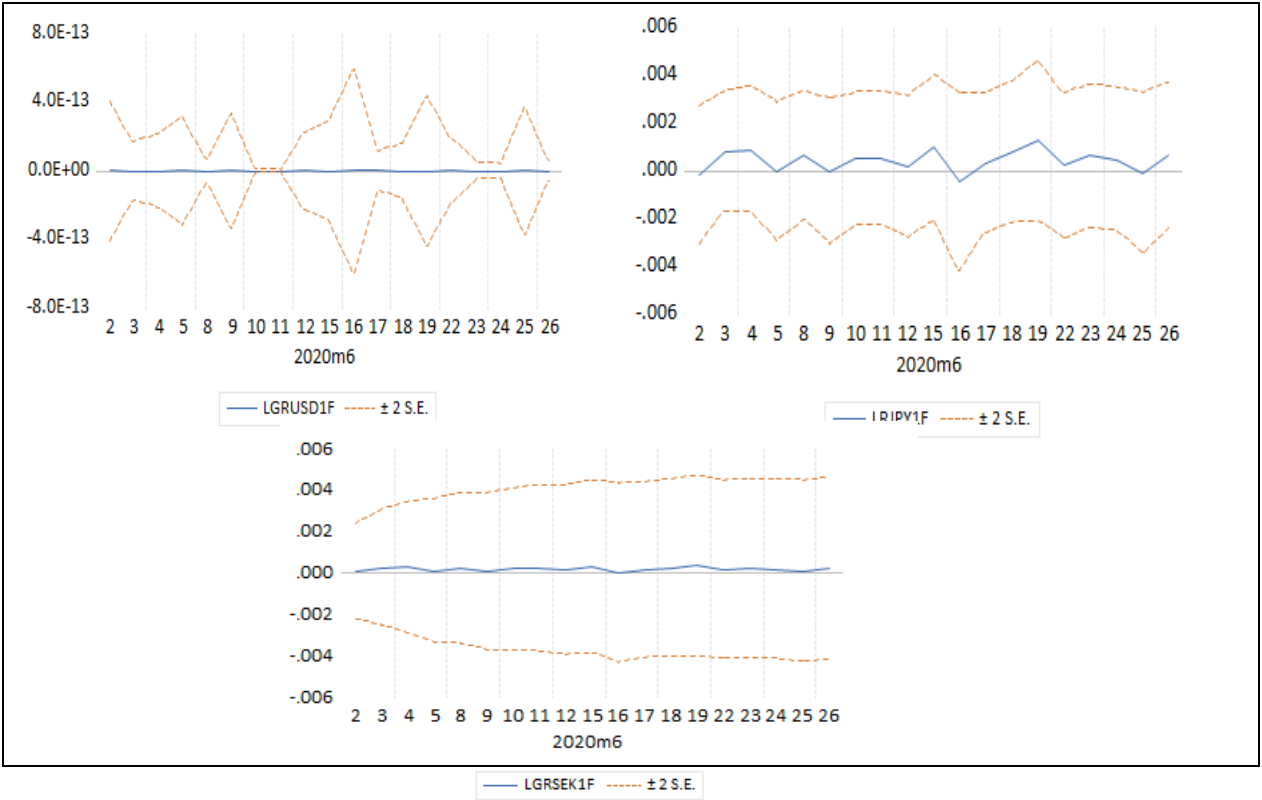

The volatility of exchange rate of BDT for the period from June 2,2020 to June 26,2020 is measured through GARCH model. The result is graphically shown in figure 6. LOGRUSDIF stands for forecasting of BDT/USD exchange rate, LGJPYIF stands for forecasting of BDT/JPY exchange rate and LGRSEKIF stands for forecasting of BDT/SEK exchange rate. Original sample of this study covers the period from March 09, 2020 to June 26, 2020. In sample volatility is measured for the period of June 2,2020 to June 26,2020.

3.6.3 Forecasting exchange rate volatility for period 2 (during COVID)

In this part, forecasting of future exchange rate volatility is done for period 2, during the pandemic time. In this part, the sample size for forecasting is considered from June 29, 2020 to December 01, 2020 and the forecasting is done for the period December 02, 2020 to December 30, 2020

Table 8: Forecast summery statistics for exchange rate volatility of BDT for Phase 2 (During COVID period)

| Volatility of BDT/USD | RMSE | 2.28E-09 |
|---|---|---|
| | MAE | 2.24E-09 |
| | Theil Inequality Coefficient | 1.00 |
| Volatility of BDT/JPY | RMSE | 0.002 |
| | MAE | 0.001 |
| | Theil Inequality Coefficient | 0.90 |
| Volatility of BDT/SEK | RMSE | 0.006 |
| | MAE | 0.004 |
| | Theil Inequality Coefficient | 0.89 |

Results in table 8 show that almost all the values in the volatility model using total of cases as a proxy of corona pandemic have a low MAE, RMSE and Theil Inequality Coefficient. The result of table 8 depicts that the model used in this study (equation 1 and 2) is very much suitable to for forecasting exchange rate of BDT. The graph shows that blue line is close gray line. It indicates the model forecasted the Exchange rate volatility very well as the original value is close to the forecasted value.

Figure 5: Exchange rate volatility forecasting (December 2,2020 to December 30,2020)

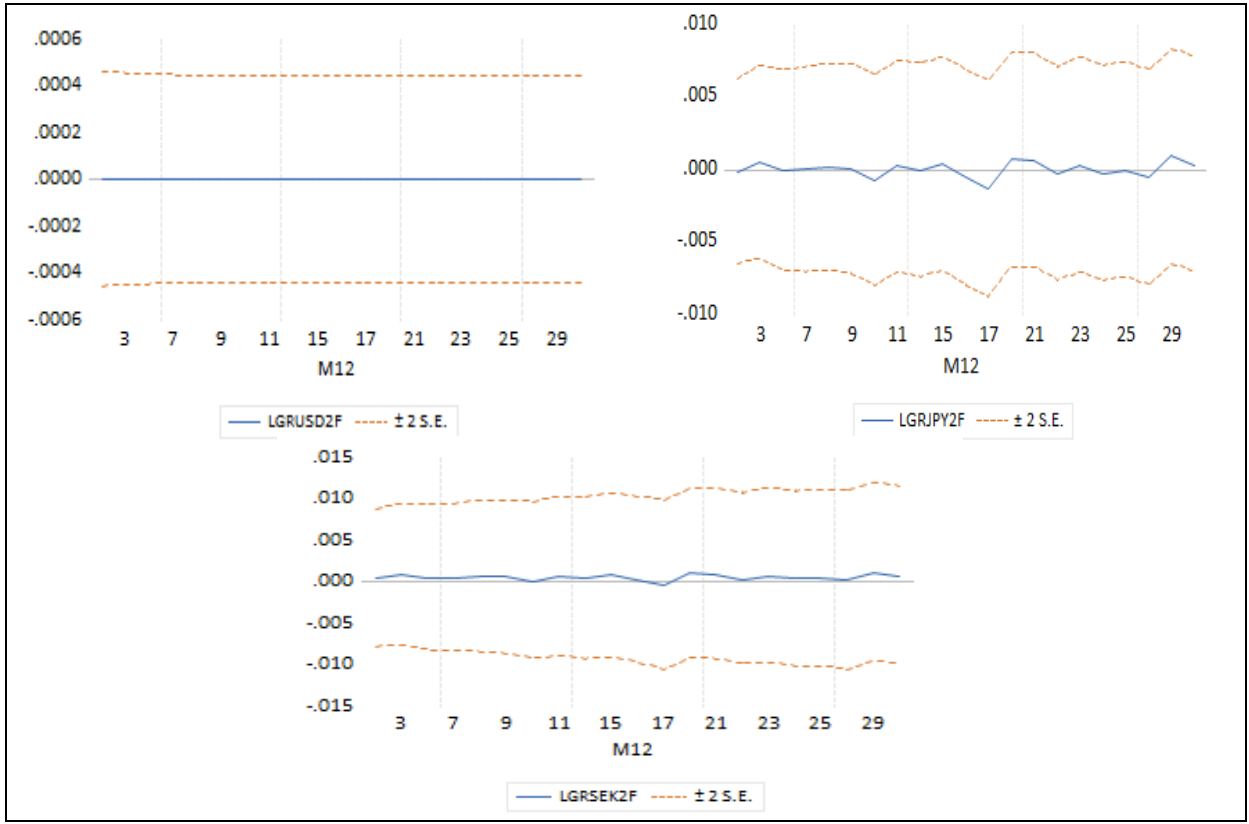

The volatility of exchange rate of BDT for the period from December 2,2020 to December 30,2020 is measured through GARCH model. The result is graphically shown in figure 6. LGRUSD2F stands for forecasting of BDT/USD exchange rate, LGRJPY2F stands for forecasting of BDT/JPY exchange rate and LGRSEK2F stands for forecasting of BDT/SEK exchange rate. The graph shows that blue line is very much aligned to gray line. It indicates the model forecasted the Exchange rate volatility very well as the original value is close to the forecasted value.

3.6.4 Comparison of forecast of the model between period1 and period 2:

The result of future exchange rate volatility of BDT for both periods, beginning of COVID and during covid are calculated. Table 8 and 9 shows the results for period 1 and period 2 respectively. For both cases the value of RMSE, MAE and Theil Inequality Coefficient are found small. But the values of parameters of period 2 is slightly smaller this those of period 1 for all the three exchange rates. For period 1, BDT/USD the value of RMSE and MAE are 0.00307 and 0.00121 on the other hand for period 2, the value of RMSE and MAE are 2.28E-09 and 2.24E-09. Similarly, for period 1, BDT/Yen the value of RMSE and MAE are 0.005 and 0.003 on the other hand for period 2, the value of RMSE and MAE are 0.002 and 0.001. Finally, Similarly, for period 1, BDT/SEK the value of RMSE and MAE are 0.007 and 0.005 on the other hand for period 2, the value of RMSE and MAE are 0.006 and 0.004.

As the less the value of RMSE and MAE the more the model is suitable for forecasting, it can be said that the model is better suited during COVID period then the beginning of COVID period.

On the same way from figure 4 and figure 5 it is found that the blue line is attached with gray line more closely during COVID period then the beginning of COVID period. Comparing Table 8 and 9 as well as figure 7 and figure 8 it can be inferred that COVID-19 variables helps to Improve the forecast of future exchange rate volatility of BDT.

# Chapter Four: Conclusion

With the outbreak of COVID-19, a large number of studies has been done on modeling and forecasting volatility in financial markets. Most of these studies focused on equity markets. Even if there have been few works measuring the volatility of exchange rate as an impact of COVID-19 neither of these studies are done on Exchange rate of BDT. This study is the first attempt to investigate the impact of Covid-19 on volatility of exchange rate return of BDT. Finding of the study are as followed

1. In Bangladesh the first case of COVID was identified in March 8, 2020. So, the study uses daily data of COVID-19 affected and daily exchange rate for 212 observations from March 09, 2020 to December 30, 2020.
2. GARCH (1,1) model is used to examine the volatility of exchange of BDT. The study found that changes in the number of COVID-19 affected cases in Bangladesh have a significant impact on volatility of return of BDT/USD, BDT/JPY and BDT/SEK.

The findings of the study will help the policy makers in foreign exchange market to take appropriate action regarding exchange rate of BDT specially during COVID-19 period.

# Reference

1. Abdullah, Siddiqua, Siddiquee, Hossain (2017), Modelling and forecasting exchange rate volatility in Bangladesh using GARCH models: a comparison based on normal and student's t-error distribution.
2. Emran, M. G. I., Raihan, M. A., Das, A. K., Islam, M. S., Khan, A.-S., & Faruq, A. T. M. O. (2023). Study on the Aftermath of Natural Ways to Cure COVID-19 in Bangladesh. International Journal of Current Science Research and Review, 6(8), 48. https://doi.org/10.47191/ijcsrr/V6-i8-48
3. Albulescu, C. 2020. Coronavirus and oil price crash: A note. arXiv preprint arXiv:2003.06184
4. ALBULESCU, C. T. (2020), "Coronavirus and financial volatility: 40 days of fasting and fear", SSRN Electronic Journal.
5. AL-AWADHI, A.M., ALSAIFI, K. AL-AWADHI, A. & ALHAMMADI, S. (2020), "Death and contagious infectious diseases: Impact of the COVID-19 virus on stock market returns", Journal of Behavioral and Experimental Finance
6. Bollerslev T (1986) Generalized autoregressive conditional heteroskedasticity. J Econ 31(3):307–327
7. Bollerslev T (1987) A conditionally heteroskedastic time series model for speculative prices and rates of return. Rev Econ Stat, 69(3):542–547
8. Bollerslev T, Chou RY, Kroner KF (1992) ARCH modeling in finance: a review of the theory and empirical evidence. J Econ 52(1–2):5–59
9. Bala DA, Asemota JO (2013) Exchange–rates volatility in Nigeria: application of GARCH models with exogenous break. CBN J Appl Stat, 4(1):89–116
10. Clement A, Samuel A (2011) Empirical modeling of Nigerian exchange rate volatility. Math Theory Model, Vol.1, No.3
11. Choo WC, Loo SC, Ahmad MI (2002) Modelling the volatility of currency exchange rate using GARCH model. Pertanika J Soc Sci Humanit 10(2):85–95

12. Engle, R.F. 1982. Autoregressive conditional heteroscedasticity with estimates of the variance of United Kingdom inflation. Econometrica: Journal of the Econometric Society, 987-1007
13. Anjom, W., & Faruq, A. T. M. O. (2023). Financial stability analysis of Islamic banks in Bangladesh. European Journal of Business and Management Research, 8(3), Article 3. https://doi.org/10.24018/ejbmr.2023.8.3.1953
14. GOODELL, J. W. (2020), "COVID-19 and finance: Agendas for future research", Finance Research Letters
15. ONALI, E. (2020), "Covid-19 and stock market volatility", SSRN Electronic Journal
16. Rofael D, Hosni R (2015) Modeling exchange rate dynamics in Egypt: observed and unobserved volatility. Mod Econ 6(01):65
17. Sadorsky, P. 2006. Modelling and forecasting petroleum futures volatility. Energy Economics, 28(4), 467-488
18. Trück, S. 2020. Modelling and forecasting volatility in the gold market. International Journal of Banking and Finance, 9(1), 48-80
19. Zeren, F., Hizarci, A. 2020. The Impact of COVID-19 Coronavirus on Stock Markets: Evidence from Selected Countries. Muhasebe ve Finans İncelemeleri Dergisi, 3(1), 78-84
20. ZHANG, D., HU, M. & JI, Q., (2020), "Financial Markets under the Global Pandemic of COVID-19", Finance Research Letters, in press
21. Bangladesh Bank website. https://www.bb.org.bd/en/index.php/econdata/exchangerate.
22. COVID-19 tracker website of Bangladesh. http://covid19tracker.gov.bd/
23. https://www.thefinancialexpress.com.bd/economy/bangladesh/bangladeshs-inflation-rises-to-644pc-in-october-highest-in-seven-years-1604550053
24. https://www.dhakatribune.com/opinion/op-ed/2020/07/29/op-ed-how-covid-19-has-impacted-the-bangladesh-economy
25. https://www.eastasiaforum.org/2020/11/07/the-shadow-of-covid-19-lingers-over-bangladeshs-economy/
26. https://www.vosesoftware.com/riskwiki/Thielinequalitycoefficient.php